# Exploring yttrium doped $C_{24}$ fullerene as a high-capacity reversible hydrogen storage material: DFT investigations

*Vikram Mahamiya[a], Alok Shukla[a*], Brahmananda Chakraborty[b,c*],*

[a]Indian Institute of Technology Bombay, Mumbai 400076, India

[b]High pressure and Synchrotron Radiation Physics Division, Bhabha Atomic Research Centre, Bombay, Mumbai, India-40085

[c]Homi Bhabha National Institute, Mumbai, India-400094

email: shukla@phy.iitb.ac.in ; brahma@barc.gov.in

## Abstract

By employing the state-of-the-art density functional theory, we report the hydrogen storage capability of yttrium decorated $C_{24}$ fullerene. Single Y atom attached on $C_{24}$ fullerene can reversibly adsorb a maximum number of 6 $H_2$ molecules with average adsorption energy -0.37 eV and average desorption temperature 477 K, suitable for fuel cell applications. The gravimetric weight content of hydrogen is 8.84 %, which exceeds the target value of 6.5 wt % H by the department of energy (DoE) of the United States. Y atom is strongly bonded to $C_{24}$ fullerene with a binding energy of -3.4 eV due to a charge transfer from Y-4d and Y-5s orbitals to the C-2p orbitals of $C_{24}$ fullerene. The interaction of $H_2$ molecules with Y atom is due to the Kubas type interaction involving a charge donation from the metal d orbital to H 1s orbital, and back donation causing slight elongation of H-H bond length. The stability of the system at the highest desorption temperature is confirmed by *ab-initio* molecular dynamics simulations, and the metal-metal clustering formation has been investigated by computing the diffusion energy barrier for the movement of Y atoms. We have corrected all the calculated energies for the van der Waals (vdW) interactions by applying the dispersion energy corrections, in addition to the contribution of the GGA exchange-correlation functional. The $C_{24}$+Y system is stable at room temperature, and at the highest desorption temperature, the presence of a sufficient diffusion energy barrier prevents metal-metal clustering. Furthermore, binding energies of $H_2$ are within the target value by DoE (-0.2-0.7 eV/$H_2$), while $H_2$ uptake (8.84 % H) is higher than DoE's criteria. Therefore, we propose that Y decorated $C_{24}$ fullerene can be tailored as a practically viable potential hydrogen storage candidate.

**Graphical abstract**

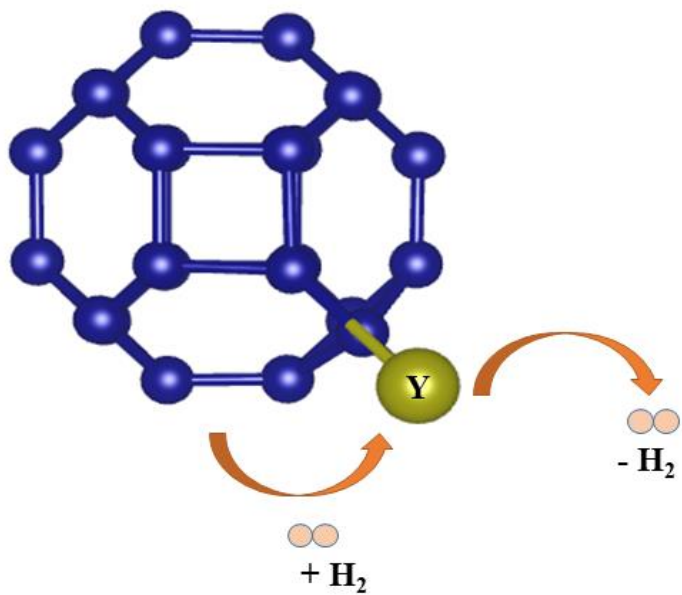

# 1. Introduction

Energy demand is increasing day by day, and the well-known sources of energy, 'fossil fuels' are continuously depleting with the increase in the global population[1]. Therefore, exploring alternative sources of green energy is one of the key challenges for the scientific community at present. Hydrogen is one of the most promising alternatives to fossil fuels because it contains the maximum energy per unit weight, it is naturally abundant, as well as the products of hydrogen combustion do not pollute the environment[2]. Although hydrogen seems to be one of the best alternatives to fossil fuels in the future, storing hydrogen for practical purposes is not so easy[2,3]. High pressure containers are required to store the hydrogen in the gaseous state, which involves safety issues[4]. Transportation is also one of the challenges for these tanks since they are rather bulky. Hydrogen storage in liquid form is costly due to the additional cost of liquefaction[5]. The solid-state form of hydrogen storage can be feasible provided the adsorption energy of the hydrogen molecules is in the range of -0.2 eV to -0.7 eV. In addition to that, the host structure should store more than 6.5 weight percentage (wt %) of hydrogen at a system level as per the guidelines of the Department of Energy, United States (DoE-US)[6,7]. The binding energy range is directly related to the desorption temperature for the reversible application of hydrogen as a potential fuel candidate.

A wide range of substrate materials, including metal alloys[8–12], metal hydrides[13–17], zeolites[18], metal organic frameworks (M.O.F.s)[19,20], are used for the solid-state adsorption and desorption of hydrogen. There are several issues with these substrate materials, such as high desorption temperature, low wt % of hydrogen storage, which need to be overcome for a practically feasible hydrogen adsorbing substrate. Metal alloys can be used as a substrate for hydrogen storage due to the stability of metal hydrides, but the problem is the gravimetric wt % of hydrogen for most of the metal alloys substrates does not lie in the criteria set by the DoE because of the presence of the heavy metal atoms[21,22]. The other issue is the high hydrogen desorption temperature for the stable metal hydrides, which suggests that they can not be used as a potential candidate for reversible hydrogen storage[23]. The cage-structured intra-crystalline zeolite[18] has also been studied for hydrogen storage purposes. The gravimetric wt % of such zeolite system[18] is much less than the prescribed value by the DoE-US. Metal-organic frameworks (M.O.F. s)[19,20] have also been studied for hydrogen storage applications. These materials have crystalline porous structures, due to which they can adsorb hydrogen molecules efficiently. Wong and their co-workers[19] have observed a 7.5 wt % hydrogen uptake in MOF-177 at 77K temperature, but the gravimetric wt % of hydrogen at room temperature is relatively low.

A lot of research has been carried out on carbon-nanomaterials as potential hydrogen storage substrates[3,24–37]. Pure carbon-nanomaterials are poor hydrogen adsorbers because the van der Waals (vdW) forces between carbon-nanomaterial and hydrogen molecules are very weak. Even at room temperature, the thermal energy exceeds the interaction forces between carbon-nanomaterial and hydrogen[38–43]. Metal doped carbon-nanomaterials are proven to be suitable hydrogen storage substrates at room temperature[26,44]. Yildirim et al.[33] have studied the hydrogen adsorption in Ti decorated carbon nanotubes. They have predicted that one Ti-atom can adsorb up to 4 $H_2$ molecules, and the hydrogen uptake is up to 7.1 wt %. But in this case, the first $H_2$ gets dissociated, which reduces the wt %. Also, the metal-metal clustering issues were not discussed in that work. Liu. et al.[38] have studied electrochemical hydrogen adsorption properties of $TiO_2$ particles doped multi-walled carbon nanotubes. Modak et al.[45] have studied hydrogen adsorption properties of Y, Zr, Nb, and Mo decorated single-wall carbon nanotubes. Chakraborty et al.[46] have investigated the reversible hydrogen adsorption in Y decorated single-wall carbon nanotube. They have predicted that one Y atom can adsorb up to 6 $H_2$ molecules, and that composition is stable up to 700 K. Hydrogen storage in calcium doped graphene was studied by Ataca et al.[37]. They found that one Ca-atom can adsorb up to 5 $H_2$ molecules, and the hydrogen uptake can reach up to 8.4 %. Lebon et al.[47] have studied the hydrogen adsorption in Ti atom decorated zigzag graphene nanoribbons using van der Waals corrected density functional theory. They have found that one Ti atom can adsorb up to 4 $H_2$ molecules, and the gravimetric wt % is above the DoE-US target. Yadav et al.[48] have investigated hydrogen adsorption in Zr decorated graphene system and found that each Zr atom can adsorb up to 9 $H_2$ with hydrogen uptake of 11%. The hydrogen storage properties of Ti decorated ψ-graphene was studied by Chakraborty et al.[34]. They have found that one Ti atom binds a maximum number of 9 $H_2$ molecules, and hydrogen uptake is 13.14 %, which is much higher than the DoE demand. Gangan et al.[49] have reported the hydrogen adsorption in the yttrium decorated graphyne system and found that the gravimetric hydrogen uptake is 10 %. The hydrogen intake capacity of Y doped $B_{40}$ is studied by Zhang et al.[50] with 5.8 % hydrogen uptake, but the problem is, in the fact that the releasing temperature is lower than the ambient temperature for their system. Soltani et al.[51] have studied hydrogen adsorption properties of palladium and cobalt atoms decorated $C_{24}$ fullerene but did not discuss the structural stability at higher temperatures.

Metal-doped carbon nanomaterials have also been studied experimentally for hydrogen storage applications. Pd doped multi-walled carbon nanotubes are found to be promising hydrogen storage systems with almost 6 % of hydrogen content, reported by Mehrabi et al.[52]. Gu et al.[53] have reported 5.7 % of hydrogen uptake for Al, Ni doped graphene nanocomposites at 473K. The composites of Mg with Ni and graphene like material (GLM) are found to be high hydrogen storage systems with a gravimetric wt % of more than 6.5 % by Tarasov et al.[12].

Among all the carbon-nanostructures, fullerene hydrides are the most studied materials for hydrogen storage experimentally. Tarasov et al.[54–57] have synthesized fullerene hydrides and explored their hydrogen storage capabilities experimentally. High content of hydrogen was obtained in the solid mixture of fullerene $C_{60}$ and $C_{70}$ with metals and intermetallic compounds[54]. The adsorption of deuterium on $C_{60}$ fullerene was studied experimentally by Tarasov et al.[58]. They have prepared deuterofullerene $C_{60}D_{24}$ and found that deuterium can be desorbed at 823 K from the composition. Reversible hydrogen storage properties of Li and Na doped $C_{60}$ fullerene were investigated by Teprovich et al. experimentally[59]. They have found that the Li and Na doped $C_{60}$ fullerene can store 5 % and 4 % of hydrogen uptake, and the desorption temperatures are 670º C and 420º C for Li and Na doped $C_{60}$ fullerenes, respectively.

Fullerene structures of carbon atoms interact strongly with the adsorbates due to the lack of unsaturated bonds. Although there are various experimental reports on the hydrogen adsorption and desorption properties of $C_{60}$ fullerene, hydrogen storage in $C_{24}$ fullerene has not been investigated experimentally so far to the best of the author's knowledge. This motivates us to explore the hydrogen storage capabilities of metal-decorated $C_{24}$ fullerene. We have selected the yttrium atom for decoration because the transition metal atoms can bind many hydrogen molecules through Kubas interactions[60,61]. The yttrium atom has only one d-electron, and it has been reported previously that transition metals with a small number of d-electrons can bind a large number of hydrogen molecules[45]. This is the motivation behind the selection of the yttrium atom for decoration purposes.

We know that yttrium atom decorated carbon nanostructures are promising candidates for hydrogen storage application[45,46,49,50]. Hydrogen storage properties of Y decorated $C_{24}$ fullerene have not been investigated so far. In this paper, we are presenting the bonding configuration and electronic properties of Y decorated $C_{24}$ fullerene with octahedral ($O_h$) point group symmetry for hydrogen storage applications. The $C_{24}$ fullerene structure having $O_h$ point group was experimentally synthesized by Oku et al.[62].

Theoretical simulations can provide a lot of crucial information, which helps the experimentalist to perform the experiment for practical purposes. However, these simulations depend very sensitively on the method used for the calculation and other theoretical parameters. There are many research papers in which high hydrogen storage capacity was reported using local density approximation (LDA) exchange-correlation, but the problem is that LDA overestimates the bonding strength of hydrogen, so hydrogen storage capacity will be lower than the reported value for the practical application[63]. The generalized gradient approximation (GGA) underestimates the bonding strength of hydrogen. As there are weak van der Waals interactions between the metal-decorated carbon nanostructures and hydrogen molecules, dispersion corrections should be accounted for in the calculations. We have used the GGA exchange-correlation functional for the calculations along with Grimme's dispersion corrections[64]. Hence, we believe that our predicted value for hydrogen uptake can be remarkably close to the experimental value. In our study, we have also calculated the diffusion energy barrier of the Y atom corresponding to its displacement from one stable site to the nearest stable site. This is also an important aspect of our work that is missing in most of the previous theoretical studies. We have also performed the *ab-initio* molecular dynamics (A.I.M.D.) simulations and found that our system is stable even at the desorption temperature of 500 K.

Here we report the hydrogen storage capability of Y decorated $C_{24}$ fullerene using density functional theory simulations. We have observed that Y atom attached on $C_{24}$ fullerene can adsorb up to 6 $H_2$ molecules with an average adsorption energy of -0.37 eV. The gravimetric hydrogen wt % for our system is 8.84 % which is much higher than the DoE-US target, and the average desorption temperature is 477 K. We have presented the electronic density of states and partial density of states analysis corresponding to the orbital interaction of Y atom on $C_{24}$ fullerene as well as for the interaction of hydrogen on $C_{24}$ + Y system. We have also calculated the Bader charge which gets transferred from Y atom $C_{24}$ fullerene, and $H_2$ molecule. *Ab-initio* molecular dynamics simulations verify the structural integrity of the metal decorated $C_{24}$ fullerene composition at higher temperatures. We hope that our theoretical work will inspire the experimentalist to tailor $C_{24}$ + Y system as high-capacity hydrogen storage material.

## 2. Computational details

All the calculations have been carried out using density functional theory (DFT) and *ab-initio* molecular dynamics (AIMD) methods. We have used Vienna *ab-initio* simulation package (VASP)[65–68] for the calculations. Since the local density approximation (LDA)[69] overestimates the binding in the system, so we have used the generalized gradient approximation (GGA) exchange-correlation functional along with the Grimme's DFT-D2[70] van der Waals corrections, to incorporate the effect of weak van der Waals interactions present in the system. The electron-correlation effects of C, Y, and H atoms are considered by using GGA exchange-correlation. In GGA potentials, the $2s^2 2p^2$, $4s^2 4p^6 4d^1 5s^2$, and $1s^1$ electronic states are considered for the valence electronic configurations for C, Y, and H atoms, respectively. We have performed the spin-polarized calculations to include the magnetic effects of the Y atom. A single unit cell of $C_{24}$ fullerene molecule was considered in a pretty large cubic box of side 30 Å to avoid the periodic interaction with other molecules. Plane-wave cut-off energy of 500 eV was used in our calculations. A Monkhorst-pack 1*1*1 K-point grid in gamma space was selected to sample the Brillouin zone. The convergence limit for the force and energy are taken as 0.01 eV/ Å and $10^{-5}$ eV for structural relaxation. We have also carried out the *ab-initio* molecular dynamics simulations to check the system's stability at desorption temperature. *Ab-initio* molecular dynamics simulations[71] were performed for Y decorated $C_{24}$ fullerene system in microcanonical (NVE) and canonical (NVT) ensemble for 5 picoseconds time duration having time step of 1 femtosecond.

## 3. Results and discussion

### 3.1 Interaction of yttrium on $C_{24}$ fullerene

One molecule of $C_{24}$ fullerene with $O_h$ point group symmetry was taken for the calculations. The DFT GGA relaxed geometry of the $C_{24}$ fullerene is shown in **Fig. 1(a).** The optimized structure of $C_{24}$ fullerene has two different kinds of bond lengths, one of which corresponds to a bond in between the common side of a hexagon and a tetragon $l_{6,4}$ = 1.49 Å, and the other one is in between the interfacial side of two hexagons of $C_{24}$ fullerene $l_{6,6}$ = 1.38 Å, which are in an excellent agreement with the literature values of $l_{6,6}$ = 1.38 Å and $l_{6,4}$ = 1.50 Å[72].

We have kept the Y atom in front of the $C_{24}$ fullerene molecule at different geometrical positions and observed two stable structures of Y-decorated $C_{24}$ fullerene as shown in **Fig. 1(b)** and **Fig. 1(c)**.

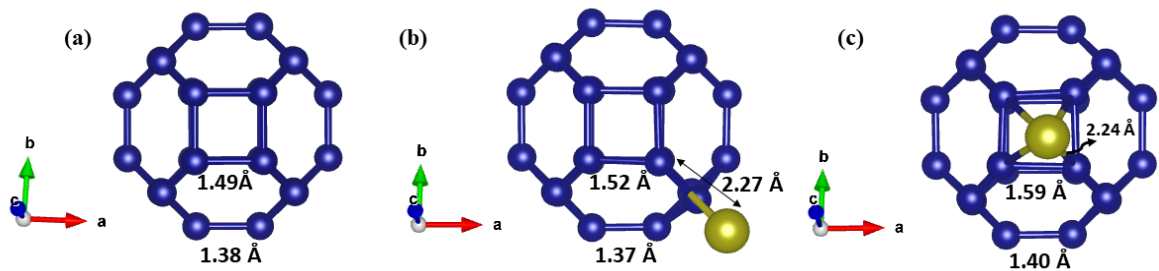

**Fig. 1 Optimized structures of (a) $C_{24}$ fullerene (b) Y-decorated $C_{24}$ fullerene where Y-atom is placed in front of the interfacial side of two hexagons (c) Y-decorated**

**C$_{24}$ fullerene where Y-atom is placed on the top of the tetragon of C$_{24}$ fullerene. Blue and golden colors correspond to C-atom and Y-atom, respectively.**

The binding energy of the Y-atom on C$_{24}$ fullerene is calculated using the following expression:

$$E_b(Y) = E(C_{24} + Y) - E(C_{24}) - E(Y) \tag{1}$$

Where $E(C_{24} + Y)$, $E(C_{24})$, and $E(Y)$ denotes the energy of Y-attached C$_{24}$ fullerene, C$_{24}$ fullerene, and single Y-atom, respectively. In **Fig. 1(b)**, Y-atom is bonded in front of the interfacial side of two hexagons at 2.27 Å distance away from C$_{24}$ fullerene with binding energy -3.40 eV, while in **Fig. 1(c)**, Y atom is bonded at 2.24 Å distance above the tetragon of C$_{24}$ fullerene with binding energy -3.07 eV. Negative binding energy indicates exothermic reactions, and since the binding energy of the Y-atom is more negative when it is attached in front of the common side of two hexagons, this structure is more stable. Since the cohesive energy of the solid bulk phase of the Y (4.42 eV/atom) is more than the binding energy of the Y atom on C$_{24}$ fullerene (3.40 eV), we have estimated the diffusion energy barrier for the Y atoms to check if there is an energy barrier for the displacements of the Y atoms, which can avoid the possibilities of metal clustering[46]. We have also performed the *ab-initio* molecular dynamics simulations to investigate whether the Y atoms remain attached to the C$_{24}$ fullerene at high releasing temperatures of hydrogen.

The binding energy of the Y atom when it is attached in front of the interfacial side of two hexagons of C$_{24}$ fullerene is more than the previously reported binding energy values of the Y atom of -2.20 eV by Chakraborty et al.[46] when it is attached to a single-wall carbon nanotube (SWCNT) and -3.15 eV by Gangan et al.[73] for Y-decorated graphyne system. As a result, this structure is more stable, and we have considered it the hydrogen adsorber substrate in our simulations. There are negligible changes in the bond lengths $l_{6,4}$ and $l_{6,6}$ of the C$_{24}$ fullerene structure after the attachment of the Y-atom.

We have plotted the density of states of $C_{24}$ fullerene and Y-attached $C_{24}$ fullerene systems in **Fig. 2** to understand their electronic structure.

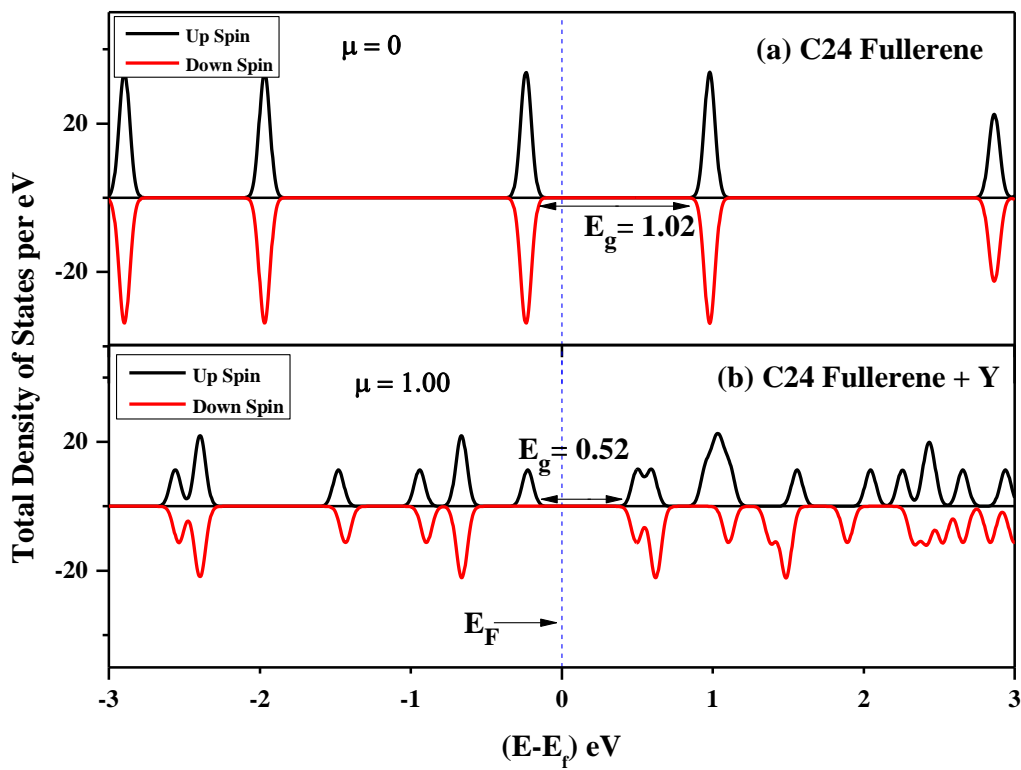

**Fig. 2 Total density of states of (a) $C_{24}$ fullerene (b) Y-decorated $C_{24}$ fullerene. μ denotes the induced magnetic moment. $E_g$ represents the energy gap in eV. Fermi level ($E_F$) is set at 0 eV.**

The density of states of $C_{24}$ fullerene is symmetric for the up and down spin, which indicates the non-magnetic characteristics of $C_{24}$ fullerene as displayed in the upper panel of **Fig. 2**. When Y-atom is attached to the $C_{24}$ fullerene, the density of states does not remain symmetric due to the induced magnetic moment in the system. The Y decorated $C_{24}$ fullerene system has a magnetic moment of 1.00 $\mu_B$ due to the 4d electron of the Y atom. The energy gap (HOMO-LUMO gap) of $C_{24}$ fullerene computed with the GGA exchange-correlation functional is 1.02 eV, which is lower than the previously reported value by Xu et al.[74] (1.79 eV using DFT – B3LYP). GGA exchange-correlation functional underestimates the energy gap of the system, so we have performed the density functional theory simulations using more sophisticated hybrid functional HSE06[75] to calculate the energy gap of $C_{24}$ fullerene. The total density of states of $C_{24}$ fullerene using hybrid functional is plotted in **Fig. 3**.

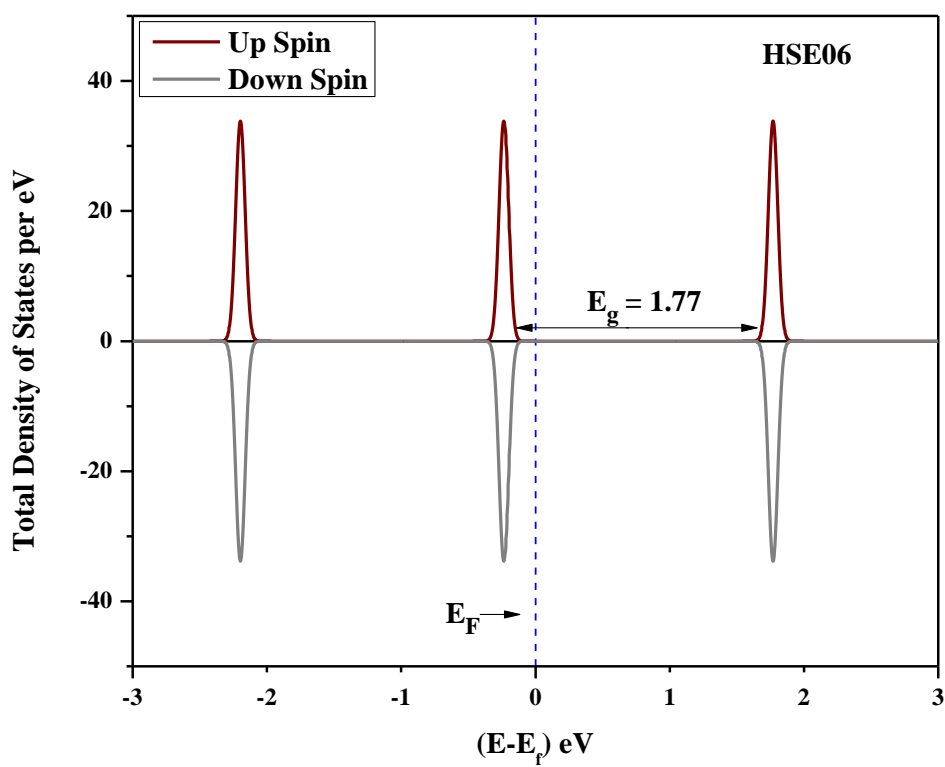

**Fig. 3 Total density of states of $C_{24}$ fullerene computed using the hybrid density function theory. $E_g$ represents the energy gap in eV. Fermi level ($E_F$) is set at 0 eV.**

We have found that the energy gap of $C_{24}$ fullerene using HSE06 functional is 1.77 eV, which is in excellent agreement with the previously reported value[74]. The energy gap of $C_{24}$ fullerene is reduced to 0.52 eV and the magnetic moment (1 μB) appears after the attachment of the Y atom, which signifies that the metallicity of the $C_{24}$ fullerene structure increases when decorated with the Y atom. Due to the increase in the metallicity of the system, more hydrogen molecules can be adsorbed by the Y decorated $C_{24}$ fullerene system compared to pristine $C_{24}$ fullerene[45].

**3.2 Adsorption of hydrogen molecules on Y-attached $C_{24}$ fullerene**

Initially, we put one hydrogen molecule at 2 Å distance away from the optimized structure of Y decorated $C_{24}$ fullerene. The structural relaxation was performed using GGA exchange-correlation functional and by employing the dispersion corrections for weak van der Waals forces. We have also considered the zero-point vibrational energy and entropy corrections for the adsorbed hydrogen molecules according to the literature[76]. The hydrogen molecule stays at a distance of 2.3 Å from the Y-atom of the $C_{24}$ + Y structure after relaxation. The optimized structure of the $C_{24}$ + Y + $H_2$ system is displayed in **Fig. 4(a).** The H-H bond length in the $C_{24}$ + Y + $H_2$ structure changes from 0.75 Å to 0.79 Å after relaxation, but the molecular nature of $H_2$ stays intact. The adsorption energy of the first $H_2$ molecule is -0.31 eV which is calculated using the following expression:

$$E_b(H_2) = E(C_{24} + Y + H_2) - E(C_{24} + Y) - E(H_2) \tag{2}$$

Where $E(C_{24} + Y + H_2)$, $E(C_{24} + Y)$ and $E(H_2)$ denotes the energies of $C_{24}$ + Y + $H_2$, $C_{24}$ + Y, and $H_2$ molecules, respectively. After getting the suitable adsorption energy of the first $H_2$ molecule as guided by DoE-US, we have kept two more $H_2$ molecules to $C_{24}$ + Y + $H_2$ structure at almost 2 Å distance away from the Y-atom in a symmetric fashion, and geometrical relaxation has been performed. The optimized structure of $C_{24}$ + Y + $3H_2$ composition is shown in **Fig. 4(b).**

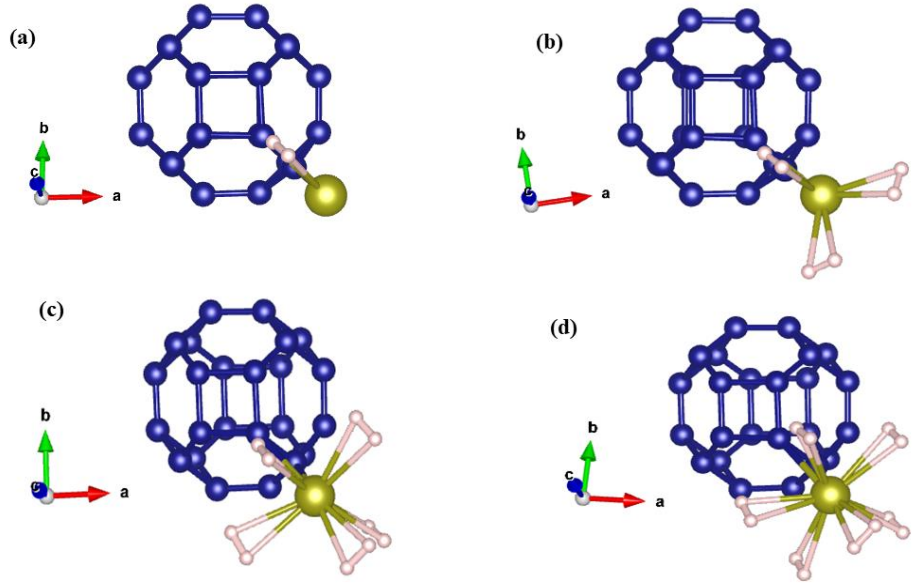

**Fig. 4** Optimized structures of (a) $C_{24}$ + Y + $H_2$ (b) $C_{24}$ + Y + $3H_2$ (c) $C_{24}$ + Y + $5H_2$ (d) $C_{24}$ + Y + $6H_2$.

The hydrogen molecules are on an average 2.3 Å distance away from the Y atom after relaxation, and the H-H bond length is 0.80 Å after relaxation. The average adsorption energy of the $2^{nd}$ and $3^{rd}$ hydrogen molecule is given by the following expression:

$$E_b(H_2) = \frac{1}{n}[E(C_{24} + Y + (m+n)H_2) - E(C_{24} + Y + m\,H_2) - nE(H_2)] \qquad (3)$$

Where $(m + n)$ is the total number of $H_2$ molecules attached to Y decorated $C_{24}$ fullerene in the current step, $m$ is the number of $H_2$ molecules attached to Y decorated $C_{24}$ fullerene system up to the previous step and $n$ is the number of $H_2$ molecules attached to Y decorated $C_{24}$ fullerene structure in the current step. The average adsorption energy of the $2^{nd}$ and $3^{rd}$ hydrogen molecule is found to be -0.39 eV which lies in the hydrogen adsorption energy range as specified by DoE-US. We have added more $H_2$ molecules subsequently in a similar fashion until the adsorption energy of hydrogen lies in the range as specified by the DoE-US. The average adsorption energy of the $4^{th}$ and $5^{th}$ $H_2$ molecules is found to be -0.38 eV. The adsorption energy of the 6$^{th}$ H$_2$ molecule is -0.37 eV. A maximum of 6 H$_2$ molecules are bonded with the Y attached C$_{24}$ fullerene with suitable adsorption energy. The average adsorption energy corresponding to all 6 H$_2$ molecules is -0.37 eV which is in the middle of the hydrogen adsorption energy range of DoE-US, hence suitable for practical adsorption and desorption process. The optimized structures of C$_{24}$ + Y + 5H$_2$ and C$_{24}$ + Y + 6H$_2$ compositions are presented in **Fig. 4(c)** and **Fig. 4(d),** respectively. The average adsorption energy of the H$_2$ molecules with GGA exchange-correlation and GGA with DFT-D2 method by taking the van der Waals interaction into account is presented in **Table 1.** The average adsorption energy of all the hydrogen molecules lies in the suitable range as described by DoE-US.

**Table 1. Average adsorption energy of Y and H$_2$ molecules with GGA and GGA+DFT-D2 method.**

| Compositions | Adsorption energy (eV) GGA | Adsorption energy (eV) GGA + DFT-D2 |
|---|---|---|
| Fullerene C$_{24}$ | 0 | 0 |
| C$_{24}$ + Y | -3.40 | - |
| C$_{24}$ + Y + H$_2$ | -0.31 | -0.36 |
| C$_{24}$ + Y + 3H$_2$ | -0.39 | -0.46 |
| C$_{24}$ + Y + 5H$_2$ | -0.38 | -0.41 |
| C$_{24}$ + Y + 6H$_2$ | -0.37 | -0.47 |
| Average adsorption | -0.37 | -0.43 |
| Energy per H$_2$ | | |
| Average desorption Temperature | 477 K | 555 K |
| Gravimetric weight % | 8.84 | |

## 3.3 Diffusion energy barrier calculations

The diffusion energy barrier of a metal atom is one of the crucial parameters for a potential hydrogen storage system. If the diffusion energy barrier of the metal is not higher than the thermal energy of the metal atom at desorption temperatures, then there is always a strong possibility of metal-metal clustering in the system. The diffusion energy barrier is the measurement of the energy barrier for the attached metal atom to diffuse from its adsorbed configurational site to the nearest site with the same kind of configuration. We have calculated the diffusion energy barrier for the Y atom using the climbing image-nudged elastic band (CI-NEB) method[77]. We have considered an in-plane horizontal path to calculate the diffusion energy barrier for the displacement of Y-atom from one stable site configuration in which Y-atom is bonded in front of the common side of two hexagons of C$_{24}$ fullerene structure to the nearest site with the same kind of configuration. Five equidistant images are considered between two optimized structures of C$_{24}$ + Y, as shown in **Fig. 5**.

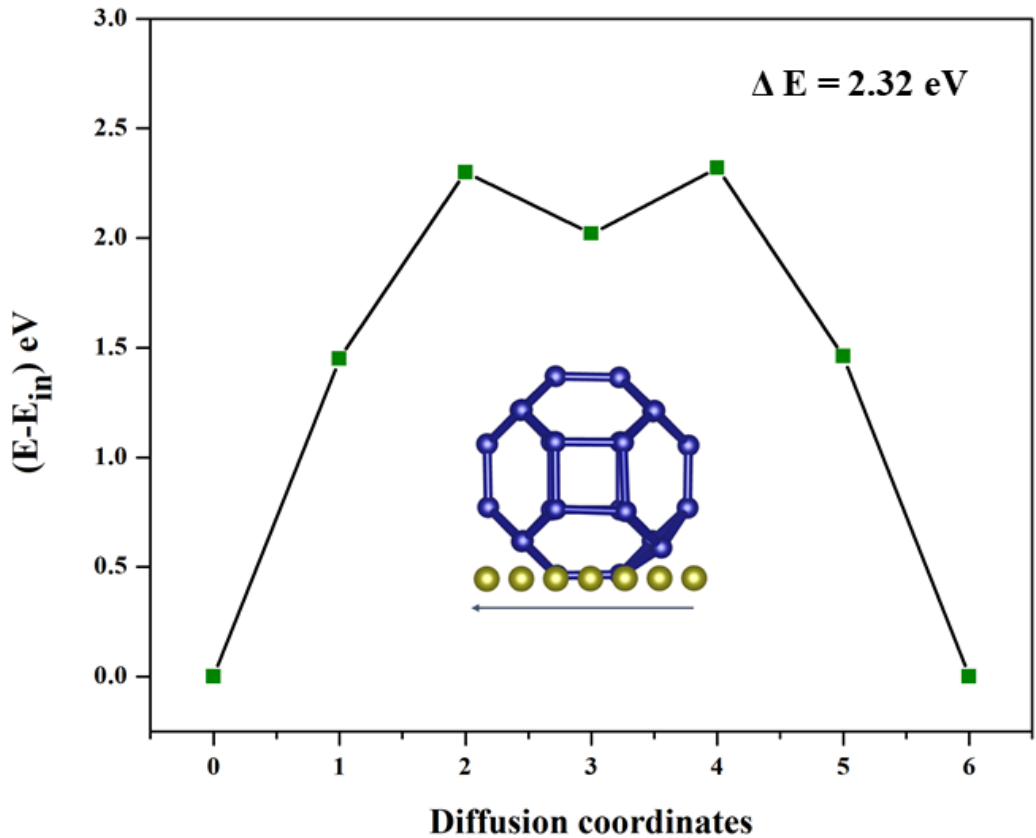

**Fig. 5 Diffusion energy barrier plot for the movement of Y-atom calculated using CI-NEB method. Energy difference of current step energy and initial energy is plotted with respect to the small displacements of Y atom.**

The diffusion energy of the Y-atom is found to be 2.32 eV which is in the range of diffusion energy of transition metals in carbon nanostructures[34,48,49]. In practical application, the quantum effects of hydrogen atoms can also affect diffusion.

## 3.4 Calculations of desorption temperature and gravimetric weight percentage of hydrogen

The desorption temperature of hydrogen is a crucial parameter of the system since it ensures that absorbed hydrogen can be used for practical purposes. We have calculated the average desorption temperature ($T_d$) of the H$_2$ molecules using Van't Hoff equation[78,79]

$$T_d = \left(\frac{E_b}{k_B}\right)\left(\frac{\Delta S}{R} - \ln P\right)^{-1} \quad (4)$$

Here $E_b$ is the average adsorption energy of the six hydrogen molecules, P is the atmospheric pressure. $k_B$, $\Delta S$, and R are Boltzmann constant, the entropy difference of hydrogen in transition from gaseous to the liquid state, and gas constant, respectively. The average adsorption energies of 6 H$_2$ molecules are -0.37 eV and -0.43 eV, corresponding to GGA and GGA + DFT-D2 method. Using these values of average adsorption energy in equation (4), we have found the values of average desorption temperature are 477 K and 555 K, corresponding to GGA and GGA + DFT-D2 method, respectively. Here we can notice that the values of average desorption temperature are quite above the room temperature and suitable for fuel cell applications.

We have loaded the Y metal atoms on C$_{24}$ fullerene such that metal-metal clustering is avoided. If we put the Y atoms on the top of the tetragons of C$_{24}$ fullerene, along with in front of the common face of the hexagons, then a very high weight percentage of hydrogen can be achieved, but there is a possibility of metal-metal clustering in such metal loading pattern. Therefore, we have loaded the metal atoms only in front of the common side of two hexagons and calculated the diffusion energy barrier for the displacement of the metal atoms. Diffusion energy barrier calculation has confirmed that the metal-metal clustering will not take place in our system. One C$_{24}$ molecule can adsorb 8 Y atoms, and each Y atom can adsorb up to 6 H$_2$ molecules, as shown in **Fig. 6.**

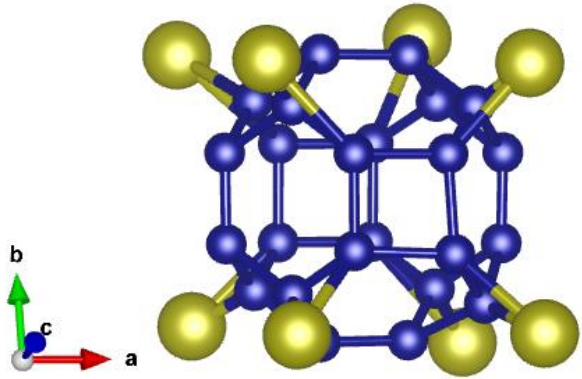

**Fig. 6 Metal loading pattern for C$_{24}$ fullerene. Weight percentage of hydrogen is 8.84 %.**

The gravimetric wt % of hydrogen is 8.84 % which is much higher than the DoE's target of 6.5 %. Experimentally up to 7.7 % of hydrogen uptake in fullerene C$_{60}$ was reported by Schur et al.[80]. We have compared the average adsorption energy of hydrogen, average desorption temperature, and gravimetric wt % H of our system with previously reported values of similar systems in **Table 2.**

**Table 2. Comparison of hydrogen storage parameters for various metal decorated carbon nanostructures.**

| Metal decorated system | Total no. of adsorbed Hydrogen molecules | Average adsorption energy per H$_2$ (eV) | Average desorption temperature (K) | Gravimetric weight percentage of H$_2$ (%) |
|---|---|---|---|---|
| Graphene + Ti[83] | 8 | -0.415 | 511.5 | 7.8 |
| Graphyne + Y[49] | 9 | -0.30 | 400 | 10 |
| SWCNT + Ti[33] | 4 | -0.18 | 230 | 8 |
| SWCNT + Y[46] | 6 | -0.41 | 524 | 6.1 |
| $B_{40}$ + Y [50] | 5 | -0.211 | 281 | 5.8 |
| $C_{24}$ + Ti[84] | 4 | -0.33 | 403.5 | 10.5 |
| *$C_{24}$ + Y* *(Present Work)* | **6** | **-0.37** | **477** | **8.84** |
| **Experimental** | | | | |
| MWCNTs + Pd[52] | - | - | - | 6.0 |
| Graphene + Ni + Al[53] | - | - | - | 5.7 |
| Graphene-Ni Nanocomposites[12] | - | - | - | >6.5 |

## 3.5 Stability of the structure at desorption temperature

The density functional theory simulations are carried out at 0 K, but for a practical hydrogen storage system, we need to check out the structural integrity at the desorption temperature. Metal atoms should be adsorbed on carbon nanostructure at ambient temperature. They should continue to be attached with the carbon nanostructure at elevated temperatures for the hydrogen desorption to take place, required for the practical use of hydrogen. We have performed *ab-* *initio* molecular dynamics (A.I.M.D.) simulations to investigate the stability of the Y-decorated $C_{24}$ fullerene structure at high temperatures. The A.I.M.D. simulations are performed in two steps, in the first step, the $C_{24}$ + Y system was placed in a microcanonical ensemble (NVE) for five picoseconds (ps), and the temperature was slowly increased up to 500 K in the time step of one femtosecond (fs). In the second step, we checked the structural solidity at 500 K temperature by keeping the system in canonical ensemble (NVT) for another five ps. We have displayed the *ab-initio* molecular dynamics simulation snapshot of the equilibrium structure of $C_{24}$ + Y after putting the system in the NVT ensemble for five ps in **Fig. 7(a).** The time evolution of the Y-C nearest neighbor distance at 500 K temperature is plotted in **Fig. 7(b).**

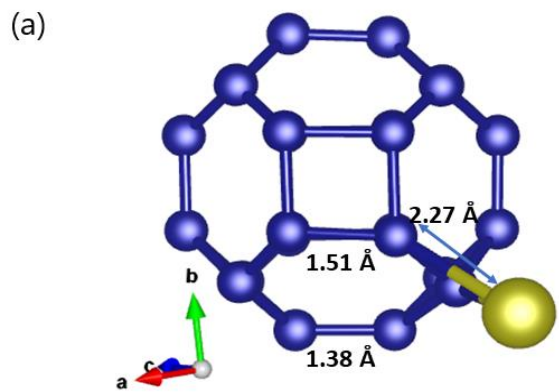

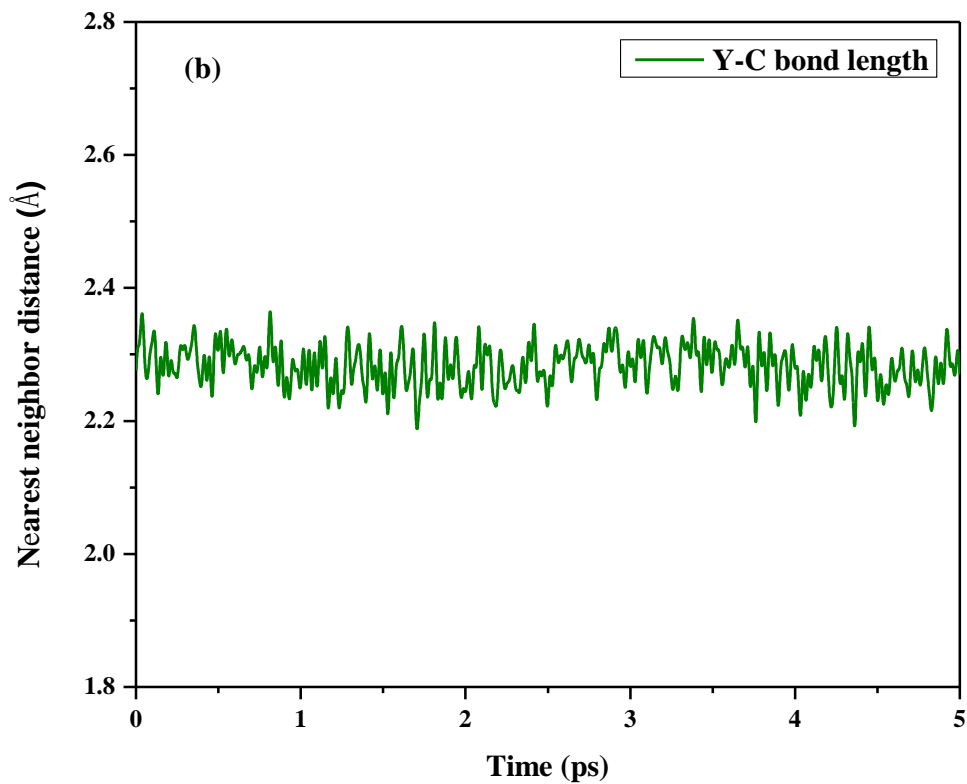

**Fig. 7(a) The *ab-initio* molecular dynamics snapshot of $C_{24}$ + Y system after putting the system in a canonical ensemble at 500 K for 5 ps. (b) Time evolution molecular dynamics picture of Y-C bond length at 500 K. The Y-decorated $C_{24}$ fullerene structure is stable at 500 K with the negligible change in C-C and C-Y bond lengths.**

We can notice that the fluctuations are very small, and the Y atom continues to be bonded. It was observed that the movement of the Y atom is negligible in $C_{24}$ + Y structure, and the change in C-Y and C-C bond length is also minor even at 500 K. These simulations ensure that the metal atom will not dislocate from the carbon nanostructure even at the desorption temperature.

### 3.6 Bonding mechanism and orbital interactions between yttrium atom and $C_{24}$ fullerene

We know that the Y-decorated $C_{24}$ fullerene can adsorb maximum of 6 $H_2$ molecules with practically suitable binding energy. The desorption temperature, diffusion energy barrier, and stability of the $C_{24}$ + Y structure at the desorption temperature suggest that our system is practically viable for hydrogen storage. To understand the orbital interaction, charge transfer mechanism, and boding mechanism, we have performed the analysis of the density of states, the partial density of states, Bader charge[81], and spatial charge density differences, discussed next.

### 3.6.1 Partial density of states (PDOS) analysis

Partial density of states provides a clear picture of orbital interactions and qualitative charge transfer behavior. We have plotted the partial density of states of C-2p orbital for the pristine $C_{24}$ fullerene and the Y decorated $C_{24}$ fullerene in **Fig. 8(a&b).** We can notice some enhancement in the partial density of states of C-2p orbital for Y-decorated $C_{24}$ fullerene in the valence band compared to that of the pristine case, which suggests that some charge has been transferred from Y-atom to the C-atoms of $C_{24}$ fullerene when Y-atom is attached. To get a better understanding of the charge transfer, we have plotted the partial density of states of 4d orbital of Y atom for isolated Y and for Y decorated $C_{24}$ fullerene in **Fig. 8(c)** and **Fig. 8(d),** respectively.

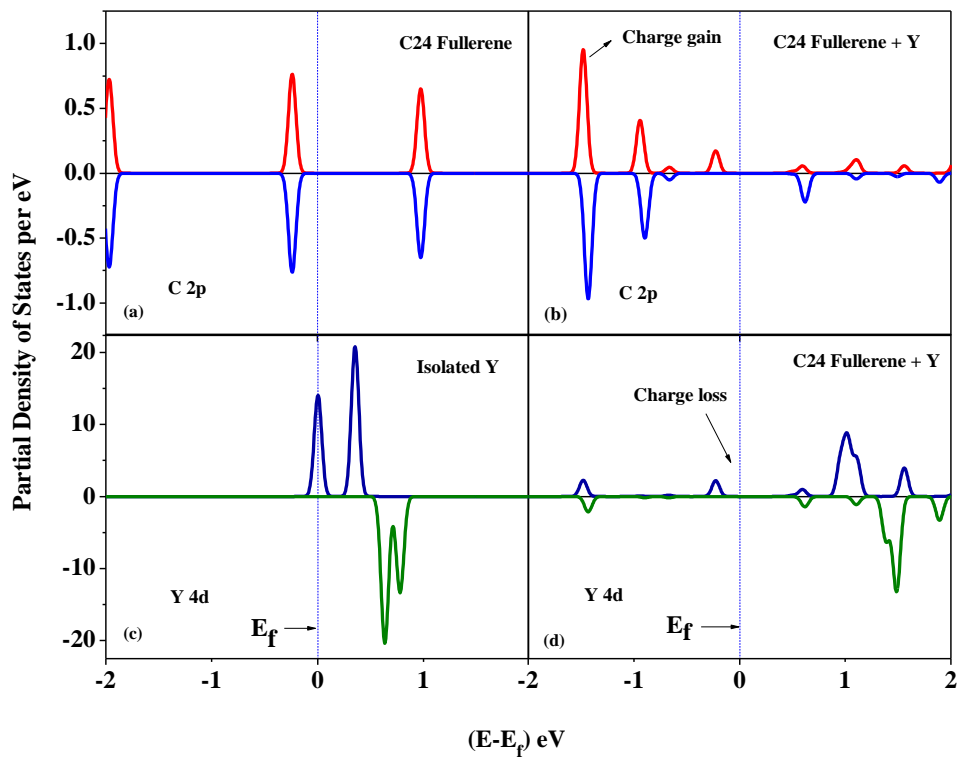

**Fig. 8** Partial density of states for (a) C-2p orbital of $C_{24}$ fullerene. (b) C-2p orbital of $C_{24}$ + Y. (c) Y-4d orbital of isolated Y atom. (d) Y-4d orbital of $C_{24}$ + Y. Fermi level is set at zero energy value.

We can see the depletion of states at the Fermi level and near Fermi level for Y-decorated $C_{24}$ fullerene compared to that of the isolated Y. That indicates the transfer of charge from Y-4d orbital to C-2p orbital when Y atom is attached to $C_{24}$ fullerene. We have also plotted the partial density of states of Y-5s orbital corresponding to isolated Y and Y-decorated $C_{24}$ fullerene system in **Fig. S1.** It is clearly visible from **Fig. S1** that there is some loss in states of Y-5s orbital in the valence band below Fermi level when Y-atom is attached to the $C_{24}$ fullerene, which indicates that some charge has been transferred from 5s orbital of Y atom to C-2p orbitals of $C_{24}$ fullerene. This charge transfer from Y 4d and 5s orbitals to C-2p orbitals of $C_{24}$ fullerene is responsible for the strong bonding between the Y atom and $C_{24}$ fullerene in $C_{24}$ + Y system.

To see the sub-orbital contributions, we have also plotted the partial density of states Y 4d-sub-orbitals ($d_{xy}$, $d_{yz}$, $d_{xz}$, $d_{x^2-y^2}$, $d_{z^2}$) for the isolated Y atom and also for the Y-decorated $C_{24}$ fullerene system in **Fig. S2.** In **Fig. S2(a).** We can see that for the isolated Y atom, there are some states at the Fermi level in the $d_{xz}$ and $d_{z^2}$ sub-orbitals which are not present in the valance band of $d_{xz}$ and $d_{z^2}$ for the $C_{24}$ + Y system (**Fig. S2(b)**). Some of the states appear in the valence band of $d_{xy}$ and $d_{yz}$ sub-orbitals for the $C_{24}$ + Y system, which was not present in isolated Y-atom. We can conclude that there is some charge redistribution among the d sub-orbitals when Y atom is attached to the $C_{24}$ fullerene.

### 3.6.2 Bader charge analysis

The density of states and partial density of states provide a qualitative picture of orbital interactions and charge transfer. To understand the charge transfer phenomenon quantitatively, we have performed the Bader charge analysis[81]. We have carried out the Bader charge calculations for the $C_{24}$ fullerene and Y decorated $C_{24}$ fullerene system. It was found that 1.13e charge transfers from the Y atom (4d and 5s orbital) to the C-atoms (2p orbitals) of $C_{24}$ fullerene. Due to this charge transfer Y atom is attached strongly to the $C_{24}$ fullerene (-3.40 eV).

### 3.6.3 Charge density difference plot

We can visualize the charge transfer phenomenon by plotting the spatial distribution of charge difference. The charge density difference 2D plot corresponding to $\rho(C_{24}$ fullerene + Y$)- \rho(C_{24}$ fullerene) is shown in **Fig. 9(a).** The charge density difference plot is for the B-G-R color pattern, and the value of iso-value is 0.106e. In **Fig. 9(a)** red color corresponds to the charge loss region, and the green, and blue color corresponds to the charge gain region. We found that some charge gets transferred from the Y atom to the C-atoms of $C_{24}$ fullerene, hence the charge density difference plot, Bader charge analysis, and partial density of states analysis are consistent.

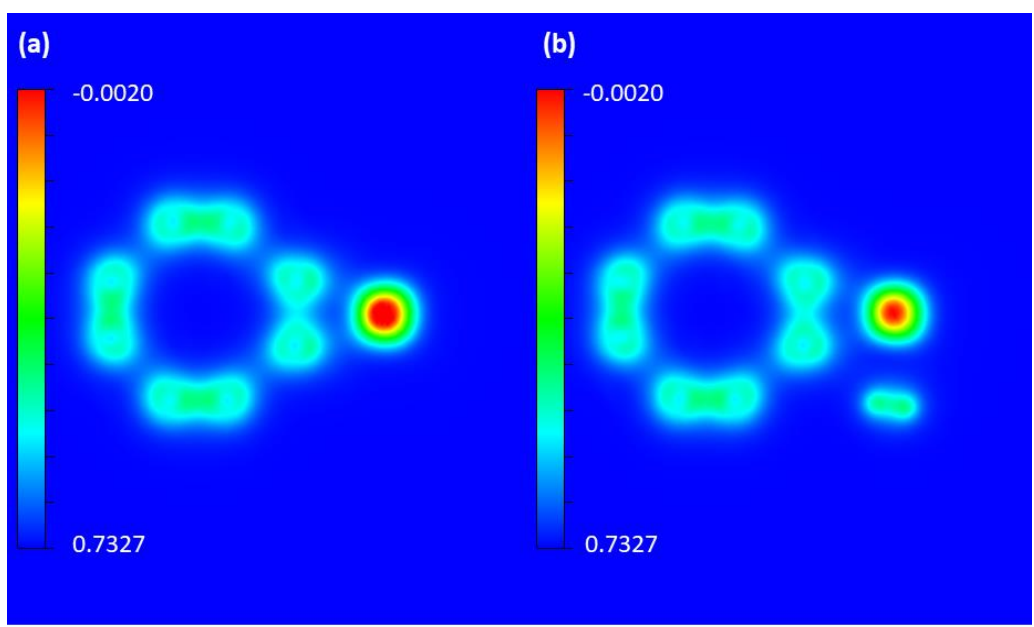

**Fig. 9 Electronic charge density difference 2D plots for (a) ρ(C$_{24}$ + Y) - ρ(C$_{24}$) system for iso-value 0.106e. (b) ρ(C$_{24}$ + Y + H$_2$)– ρ(C$_{24}$ + Y) system for iso-value 0.162e. Here red color denotes charge loss region and green and blue color denotes less and more charge gain regions, respectively.**

## 3.7 Bonding mechanism and orbital interactions between H$_2$ molecules and Y-atom attached with C$_{24}$ fullerene

In this section, we will explain the nature of the interaction, more specifically, the type of bonding between hydrogen and Y-decorated C$_{24}$ fullerene.

### 3.7.1 Kubas-type interaction

We have found that the average adsorption energy of $H_2$ molecules on Y-decorated $C_{24}$ fullerene is -0.37 eV, which is significantly higher than the adsorption energy corresponding to the physisorption process but lower than the chemisorption process. We have observed that the H-H bond length shows slight elongation when $H_2$ molecules get attached to the Y atom of the $C_{24}$ + Y system. The H-H bond length changes from 0.75 Å to 0.79 Å in the first $H_2$ molecule when it is attached to the Y atom. The slight change in the H-H bond length suggests that it is not a chemisorption process. The interaction between the Y atom of $C_{24}$ + Y and the hydrogen molecules is in between the physisorption and chemisorption process in strength, which is called Kubas type interaction[60,61,82]. Kubas interaction involves the transfer of charge from the filled σ (HOMO) orbital of the $H_2$ molecule to the vacant 4d orbital of Y atom and subsequently some back donation of charge from the filled 4d orbital of Y atom to the empty σ* (LUMO) orbital of the $H_2$ molecule. In the whole process of charge transfer, there is some net charge gain by the s orbital of hydrogen atoms. As a result, the H-H bond length increases slightly. Hence the bonding of hydrogen molecules with the metal atom is mainly due to the Kubas type interactions and the weak van der Waals interactions.

**3.7.2 Partial density of states (PDOS) analysis**

To further understand the charge transfer and bonding mechanisms between the hydrogen molecules and the Y atom, which is attached to the $C_{24}$ fullerene, we have plotted the partial density of states (PDOS) of H 1s orbital for the isolated hydrogen molecule and $C_{24}$ + Y + $H_2$ composition. We have also plotted the PDOS of Y 4d orbital for Y-decorated $C_{24}$ fullerene and $C_{24}$ + Y + $H_2$ composition as displayed in **Fig. 10.**

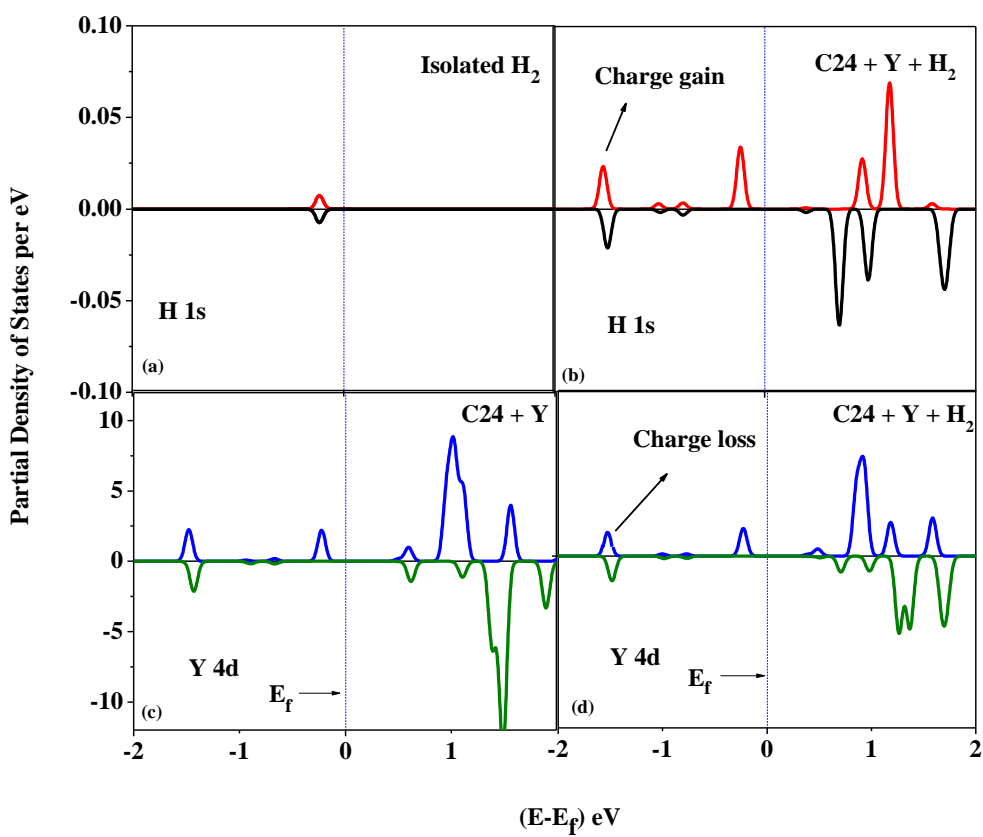

**Fig. 10** Partial density of states for (a) H 1s orbital of isolated $H_2$ molecule. (b) H 1s orbital for $C_{24}$ + Y + $H_2$. (c) Y 4d orbital of $C_{24}$ + Y. (d) Y 4d orbital of $C_{24}$ + Y + $H_2$. Fermi energy is set at zero energy value.

From **Fig. 10 (a)** and **Fig. 10 (b),** we can conclude that some charge has been transferred to H 1s orbital when hydrogen molecule is attached to Y decorated $C_{24}$ fullerene since enhancement in the partial density of states of H 1s orbital in the valance band is clearly visible in **Fig. 10 (b).** In **Fig. 10 (c),** we plotted the partial density of states of 4d orbital of Y atom when Y atom is attached to $C_{24}$ fullerene, while in **Fig. 10 (d),** we plotted the PDOS of Y-4d orbital for $C_{24}$ + Y + $H_2$ composition. The partial density of states of Y 4d orbital shows some small change in the valence band as shown in **Fig. 10 (c)** and **Fig. 10 (d),** which corresponds to the transfer of charge from Y 4d orbital to H 1s orbital.

### 3.7.3 Bader charge analysis

We have carried out the Bader charge analysis for the $C_{24}$ + Y and $C_{24}$ + Y + $H_2$ structures. It was observed that a net charge of 0.06e has been transferred from Y atom to 1s orbital of H on adsorption of the first hydrogen molecule. The small charge 0.06e, which is transferred from the Y-atom to H-atoms, is responsible for the small increase in the H-H bond length in $C_{24}$ + Y + $H_2$ system.

### 3.7.4 Charge density difference plot

The 2d charge density difference plot corresponding to $\rho(C_{24} + Y + H_2) - \rho(C_{24} + Y)$ is displayed in **Fig. 9(b).** The plot is according to the B-G-R color pattern, and the iso-value is 0.162e. The red color is for the charge loss region, and the green and blue colors are for less and more charge gain regions, respectively. From the figure, it is obvious that the charge has been transferred from the Y atom to the C atoms of $C_{24}$ fullerene and the H atoms of $H_2$. We can see in **Fig. 9(b)** that the $C_{24}$ fullerene region is of green and blue color while the $H_2$ region is of green color only, which explains that more charge has been transferred from the Y atom to the C atoms of $C_{24}$ fullerene than from Y atom to H atoms of $H_2$. Clearly, our charge density difference plot is consistent with Bader charge and partial density of states analysis.

## 4 Conclusions

We have performed extensive DFT simulations to investigate the hydrogen storage capability of Y decorated $C_{24}$ fullerene. Y atom is bonded on $C_{24}$ fullerene with a binding energy of -3.4 eV due to charge transfer from Y 4d and 5s orbitals to C-2p orbitals of $C_{24}$ fullerene. Y decorated $C_{24}$ fullerene can adsorb up to 6 $H_2$ molecules per metal atom with average adsorption energy -0.37 eV, suitable for hydrogen adsorption as guided by DoE-US. The hydrogen molecules are adsorbed on the $C_{24}$+Y system by Kubas type interaction. The average desorption temperature is 477 K which is well above the room temperature and lies in the suitable range for fuel cell application. To understand the metal-metal clustering issue, we have calculated the diffusion energy barrier for the movement of Y atoms, which is sufficiently large (2.32 eV). We have also performed the *ab-initio* molecular dynamics simulations to check the stability of metal decorated $C_{24}$ fullerene structure at desorption temperature and observed that the structure is stable at high temperatures. To understand the electronic structure and charge transfer mechanisms, we have analyzed the partial density of states and performed Bader charge analysis. Our calculations suggest that one $C_{24}$ molecule can adsorb 8 Y atoms leading to 8.84 weight percentage of hydrogen, which is much higher than the DoE-US target. We propose that Y decorated $C_{24}$ fullerene is a potential hydrogen storage candidate, and our results will motivate the experimentalist to check the hydrogen storage capability of Y decorated $C_{24}$ fullerene.

## Acknowledgment

VM would like to acknowledge DST-INSPIRE for providing the fellowship. BC would like to thank Dr. T. Shakuntala and Dr. Nandini Garg for support and encouragement. BC also acknowledge support from Dr. S.M. Yusuf and Dr. A. K Mohanty.

**REFERENCES:**

[1] Hoel M, Kvemdokk S. Depletion of fossil fuels and the impacts of global warming. vol. 18. 1996.

[2] Mazloomi K, Gomes C. Hydrogen as an energy carrier: Prospects and challenges. Renew Sustain Energy Rev 2012;16:3024–33. https://doi.org/10.1016/j.rser.2012.02.028.

[3] Xia Y, Yang Z, Zhu Y. Porous carbon-based materials for hydrogen storage: Advancement and challenges. J Mater Chem A 2013;1:9365–81. https://doi.org/10.1039/c3ta10583k.

[4] Sinigaglia T, Lewiski F, Santos Martins ME, Mairesse Siluk JC. Production, storage, fuel stations of hydrogen and its utilization in automotive applications-a review. Int J Hydrogen Energy 2017;42:24597–611. https://doi.org/10.1016/j.ijhydene.2017.08.063.

[5] Yang J, Sudik A, Wolverton C, Siegel DJ. High capacity hydrogen storage materials: Attributes for automotive applications and techniques for materials discovery. Chem Soc Rev 2010;39:656–75. https://doi.org/10.1039/b802882f.

[6] Ströbel R, Garche J, Moseley PT, Jörissen L, Wolf G. Hydrogen storage by carbon materials. J Power Sources 2006;159:781–801. https://doi.org/10.1016/j.jpowsour.2006.03.047.

[7] DOE technical system targets for onboard hydrogen storage for light-duty fuel cell vehicles. Https://WwwEnergyGov/ Eere/Fuelcells/Doe-Technical-Targets-Onboard-Hydrogenstorage-Light-Duty-Vehicles n.d.

[8] Yu XB, Wu Z, Xia BJ, Xu NX. Enhancement of hydrogen storage capacity of Ti-V-

Cr-Mn BCC phase alloys. J Alloys Compd 2004;372:272–7. https://doi.org/10.1016/j.jallcom.2003.09.153.

[9] Zaluski L, Zaluska A, Str J, Schulz R. ALLOY5 AND COMPOUNDS Effects of relaxation on hydrogen absorption in Fe-Ti produced by ball-milling. vol. 227. 1995.

[10] Wijayanti ID, Denys R, Suwarno, Volodin AA, Lototskyy M V., Guzik MN, et al. Hydrides of Laves type Ti–Zr alloys with enhanced H storage capacity as advanced metal hydride battery anodes. J Alloys Compd 2020;828:154354. https://doi.org/10.1016/j.jallcom.2020.154354.

[11] Nyallang Nyamsi S, Lototskyy M V., Yartys VA, Capurso G, Davids MW, Pasupathi S. 200 NL H2 hydrogen storage tank using MgH2–TiH2–C nanocomposite as H storage material. Int J Hydrogen Energy 2021;46:19046–59. https://doi.org/10.1016/j.ijhydene.2021.03.055.

[12] Tarasov BP, Arbuzov AA, Mozhzhuhin SA, Volodin AA, Fursikov P V., Lototskyy M V., et al. Hydrogen storage behavior of magnesium catalyzed by nickel-graphene nanocomposites. Int J Hydrogen Energy 2019;44:29212–23. https://doi.org/10.1016/j.ijhydene.2019.02.033.

[13] Sakintuna B, Lamari-Darkrim F, Hirscher M. Metal hydride materials for solid hydrogen storage: A review. Int J Hydrogen Energy 2007;32:1121–40. https://doi.org/10.1016/j.ijhydene.2006.11.022.

[14] Bellosta von Colbe J, Ares JR, Barale J, Baricco M, Buckley C, Capurso G, et al. Application of hydrides in hydrogen storage and compression: Achievements, outlook and perspectives. Int J Hydrogen Energy 2019;44:7780–808. https://doi.org/10.1016/j.ijhydene.2019.01.104.

[15] Yartys VA, Harris IR, Panasyuk V V. Hydrogen in Metals New Metal Hydrides : a Survey. Energy 2001;37:69–86.

[16] Suwarno S, Lototskyy M V., Yartys VA. Thermal desorption spectroscopy studies of hydrogen desorption from rare earth metal trihydrides REH3 (RE=Dy, Ho, Er). J Alloys Compd 2020;842:155530. https://doi.org/10.1016/j.jallcom.2020.155530.

[17] Young KH, Nei J, Wan C, Denys R V., Yartys VA. Comparison of C14-and C15-predomiated AB 2 metal hydride alloys for electrochemical applications. Batteries 2017;3:1–19. https://doi.org/10.3390/batteries3030022.

[18] Kleperis J, Lesnicenoks P, Grinberga L, Chikvaidze G, Klavins J. Zeolite as material for hydrogen storage in transport applications. Latv J Phys Tech Sci 2013;50:59–64. https://doi.org/10.2478/lpts-2013-0020.

[19] Wong-Foy AG, Matzger AJ, Yaghi OM. Exceptional H2 saturation uptake in microporous metal-organic frameworks. J Am Chem Soc 2006;128:3494–5. https://doi.org/10.1021/ja058213h.

[20] Hailian Li*, Mohamed Eddaoudi[2] MO& OMY. Design and synthesis of an exceptionally stable and highly porous metal-organic framework. Nature 1999:276–9.

[21] Singh AK, Singh AK, Srivastava ON. On the synthesis of the Mg2Ni alloy by mechanical alloying. vol. 227. 1995.

[22] Floriano R, Leiva DR, Dessi JG, Asselli AAC, Junior AMJ, Botta WJ. Mg-based nanocomposites for hydrogen storage containing Ti-Cr-V alloys as additives. Mater Res 2016;19:80–5. https://doi.org/10.1590/1980-5373-MR-2016-0179.

[23] Heung LK. Using Metal Hydride to Store Hydrogen 2003:8.

[24] Pupysheva O V., Farajian AA, Yakobson BI. Fullerene nanocage capacity for hydrogen storage. Nano Lett 2008;8:767–74. https://doi.org/10.1021/nl071436g.

[25] Sankaran M, Viswanathan B. The role of heteroatoms in carbon nanotubes for hydrogen storage. Carbon N Y 2006;44:2816–21. https://doi.org/10.1016/j.carbon.2006.03.025.

[26] Cabria I, López MJ, Alonso JA. Enhancement of hydrogen physisorption on graphene and carbon nanotubes by Li doping. J Chem Phys 2005;123. https://doi.org/10.1063/1.2125727.

[27] Zhang Y, Cheng X. Hydrogen storage property of alkali and alkaline-earth metal atoms decorated C24 fullerene: A DFT study. Chem Phys 2018;505:26–33. https://doi.org/10.1016/j.chemphys.2018.03.010.

[28] Chen X, Yuan F, Gu Q, Yu X. Light metals decorated covalent triazine-based frameworks as a high capacity hydrogen storage medium. J Mater Chem A 2013;1:11705–10. https://doi.org/10.1039/c3ta11940h.

[29] Paul D, Deb J, Bhattacharya B, Sarkar U. Electronic and optical properties of C 24 , C 12 X 6 Y 6 , and X 12 Y 12 ( X = B , Al and Y = N , P ) 2018;12.

[30] Patchkovskii S, Tse JS, Yurchenko SN, Zhechkov L, Heine T, Seifert G. Graphene nanostructures as tunable storage media for molecular hydrogen. Proc Natl Acad Sci U S A 2005;102:10439–44. https://doi.org/10.1073/pnas.0501030102.

[31] Muniz AR, Singh T, Maroudas D. Effects of hydrogen chemisorption on the structure and deformation of single-walled carbon nanotubes. Appl Phys Lett 2009;94. https://doi.org/10.1063/1.3095923.

[32] Tada K, Furuya S, Watanabe K. Ab initio study of hydrogen adsorption to single-walled carbon nanotubes. Phys Rev B - Condens Matter Mater Phys 2001;63:1–3. https://doi.org/10.1103/PhysRevB.63.155405.

[33] Yildirim T, Ciraci S. Titanium-decorated carbon nanotubes as a potential high-capacity hydrogen storage medium. Phys Rev Lett 2005;94:1–4. https://doi.org/10.1103/PhysRevLett.94.175501.

[34] Chakraborty B, Ray P, Garg N, Banerjee S. High capacity reversible hydrogen storage in titanium doped 2D carbon allotrope Ψ-graphene: Density Functional Theory investigations. Int J Hydrogen Energy 2021;46:4154–67. https://doi.org/10.1016/j.ijhydene.2020.10.161.

[35] Durgun E, Ciraci S, Yildirim T. Functionalization of carbon-based nanostructures with light transition-metal atoms for hydrogen storage. Phys Rev B - Condens Matter Mater Phys 2008;77:1–9. https://doi.org/10.1103/PhysRevB.77.085405.

[36] Yildirim T, Íñiguez J, Ciraci S. Molecular and dissociative adsorption of multiple hydrogen molecules on transition metal decorated C60. Phys Rev B - Condens Matter Mater Phys 2005;72:3–6. https://doi.org/10.1103/PhysRevB.72.153403.

[37] Ataca C, Aktürk E, Ciraci S. Hydrogen storage of calcium atoms adsorbed on graphene: First-principles plane wave calculations. Phys Rev B - Condens Matter Mater Phys 2009;79:1–4. https://doi.org/10.1103/PhysRevB.79.041406.

[38] Liu E, Wang J, Li J, Shi C, He C, Du X, et al. Enhanced electrochemical hydrogen storage capacity of multi-walled carbon nanotubes by TiO2 decoration. Int J Hydrogen Energy 2011;36:6739–43. https://doi.org/10.1016/j.ijhydene.2011.02.128.

[39] Züttel A. Materials for hydrogen storage. Mater Today 2003;6:24–33.

https://doi.org/10.1016/S1369-7021(03)00922-2.

[40] Ma LP, Wu ZS, Li J, Wu ED, Ren WC, Cheng HM. Hydrogen adsorption behavior of graphene above critical temperature. Int J Hydrogen Energy 2009;34:2329–32. https://doi.org/10.1016/j.ijhydene.2008.12.079.

[41] Klangt CH, Bethunet DS, Heben MJ. letters to nature " ' 0 Iron cycle 1997;668:1995–7.

[42] Ding F, Lin Y, Krasnov PO, Yakobson BI. Nanotube-derived carbon foam for hydrogen sorption. J Chem Phys 2007;127. https://doi.org/10.1063/1.2790434.

[43] Lee SY, Park SJ. Influence of the pore size in multi-walled carbon nanotubes on the hydrogen storage behaviors. J Solid State Chem 2012;194:307–12. https://doi.org/10.1016/j.jssc.2012.05.027.

[44] Liu Y, Brown CM, Neumann DA, Geohegan DB, Puretzky AA, Rouleau CM, et al. Metal-assisted hydrogen storage on Pt-decorated single-walled carbon nanohorns. Carbon N Y 2012;50:4953–64. https://doi.org/10.1016/j.carbon.2012.06.028.

[45] Modak P, Chakraborty B, Banerjee S. Study on the electronic structure and hydrogen adsorption by transition metal decorated single wall carbon nanotubes. J Phys Condens Matter 2012;24. https://doi.org/10.1088/0953-8984/24/18/185505.

[46] Chakraborty B, Modak P, Banerjee S. Hydrogen storage in yttrium-decorated single walled carbon nanotube. J Phys Chem C 2012;116:22502–8. https://doi.org/10.1021/jp3036296.

[47] Lebon A, Carrete J, Gallego LJ, Vega A. Ti-decorated zigzag graphene nanoribbons for hydrogen storage. A van der Waals-corrected density-functional study. Int. J.

Hydrogen Energy, vol. 40, Elsevier Ltd; 2015, p. 4960–8. https://doi.org/10.1016/j.ijhydene.2014.12.134.

[48] Yadav A, Chakraborty B, Gangan A, Patel N, Press MR, Ramaniah LM. Magnetic Moment Controlling Desorption Temperature in Hydrogen Storage: A Case of Zirconium-Doped Graphene as a High Capacity Hydrogen Storage Medium. J Phys Chem C 2017;121:16721–30. https://doi.org/10.1021/acs.jpcc.7b04886.

[49] Gangan A, Chakraborty B, Ramaniah LM, Banerjee S. First principles study on hydrogen storage in yttrium doped graphyne: Role of acetylene linkage in enhancing hydrogen storage. Int J Hydrogen Energy 2019;44:16735–44. https://doi.org/10.1016/j.ijhydene.2019.05.051.

[50] Zhang Y, Han X, Cheng X. The high capacity hydrogen storage material of Y-doped B40: A theoretical study. Chem Phys Lett 2020;739:136961. https://doi.org/10.1016/j.cplett.2019.136961.

[51] Soltani A, Javan MB, Hoseininezhad-Namin MS, Tajabor N, Lemeski ET, Pourarian F. Interaction of hydrogen with Pd- and co-decorated C24 fullerenes: Density functional theory study. Synth Met 2017;234:1–8. https://doi.org/10.1016/j.synthmet.2017.10.004.

[52] Mehrabi M, Parvin P, Reyhani A MS. Hydrogen storage in multi-walled carbon nanotubes decorated with palladium nanoparticles using laser ablation/chemical reduction methods. Mater Res Express 2017;4:095030.

[53] Gu J, Zhang X, Fu L, Pang A. Study on the hydrogen storage properties of the dual active metals Ni and Al doped graphene composites. Int J Hydrogen Energy 2019;44:6036–44. https://doi.org/10.1016/j.ijhydene.2019.01.057.

[54] Tarasov BP, Fokin VN, Moravsky AP, Shul'ga YM, Yartys VA. Hydrogenation of fullerenes C60 and C70 in the presence of hydride-forming metals and intermetallic compounds. J Alloys Compd 1997;253–254:25–8. https://doi.org/10.1016/S0925-8388(96)03073-3.

[55] Schur D V., Tarasov BP, Shul'ga YM, Zaginaichenko SY, Matysina ZA, Pomytkin AP. Hydrogen in fullerites. Carbon N Y 2003;41:1331–42. https://doi.org/10.1016/S0008-6223(03)00057-5.

[56] Tarasov BP, Fursikov P V., Volodin AA, Bocharnikov MS, Shimkus YY, Kashin AM, et al. Metal hydride hydrogen storage and compression systems for energy storage technologies. Int J Hydrogen Energy 2021;46:13647–57. https://doi.org/10.1016/j.ijhydene.2020.07.085.

[57] B.P.Tarasov, V.N.Fokin, A.P.Moravsky, Yu.M.Shul'ga, V.A.Yartys DVS. No Title. Promot. Fuller. hydride Synth. by Intermet. Compd., Buenos Aires, Argentina.: Hydrogen Energy Progress XII Proceedings of the 12th Word Hydrogen Energy Conference; 1998, p. 1221–30.

[58] Tarasov BP, Shul'Ga YM, Fokin VN, Vasilets VN, Shul'Ga NY, Schur D V., et al. Deuterofullerene C60D24 studied by XRD, IR and XPS. J Alloys Compd 2001;314:296–300. https://doi.org/10.1016/S0925-8388(00)01257-3.

[59] Teprovich JA, Knight DA, Peters B, Zidan R. Comparative study of reversible hydrogen storage in alkali-doped fulleranes. J Alloys Compd 2013;580:S364–7. https://doi.org/10.1016/j.jallcom.2013.02.024.

[60] Kubas GJ. Fundamentals of H 2 binding and reactivity on transition metals underlying hydrogenase function and H 2 production and storage. Chem Rev 2007;107:4152–205.

https://doi.org/10.1021/cr050197j.

[61] Kubas GJ. Hydrogen activation on organometallic complexes and H2 production, utilization, and storage for future energy. J Organomet Chem 2009;694:2648–53. https://doi.org/10.1016/j.jorganchem.2009.05.027.

[62] Oku T, Kuno M, Kitahara H, Narita I. Formation, atomic structures and properties of boron nitride and carbon nanocage fullerene materials. Int J Inorg Mater 2001;3:597–612. https://doi.org/10.1016/S1466-6049(01)00169-6.

[63] Netrattana P, Reunchan P. First-principles study of hydrogen adsorption on two-dimensional C 2 N sheet n.d.:1–4.

[64] Grimme S. Semiempirical GGA-type density functional constructed with a long-range dispersion correction. J Comput Chem 2006;27:1787–99. https://doi.org/10.1002/jcc.20495.

[65] Kresse G, Furthmiiller B ' J. Efficiency of ab-initio total energy calculations for metals and semiconductors using a plane-wave basis set. vol. 6. 1996.

[66] Kresse G, Hafner J. Ab. initio molecular dynamics for liquid metals. vol. 47. n.d.

[67] Kresse G. Ab initio molecular-dynamics simulation of the liquid-metal-amorphous-semiconductor transition in germanium. vol. 8. n.d.

[68] Kresse G, Furthmü J. Efficient iterative schemes for ab initio total-energy calculations using a plane-wave basis set. 1996.

[69] Lugo-Solis A, Vasiliev I. Ab initio study of K adsorption on graphene and carbon nanotubes: Role of long-range ionic forces. Phys Rev B - Condens Matter Mater Phys 2007;76. https://doi.org/10.1103/PhysRevB.76.235431.

[70] Grimme S, Antony J, Ehrlich S, Krieg H. A consistent and accurate ab initio parametrization of density functional dispersion correction (DFT-D) for the 94 elements H-Pu. J Chem Phys 2010;132. https://doi.org/10.1063/1.3382344.

[71] Nosé S. A molecular dynamics method for simulations in the canonical ensemble. Mol Phys 1984;52:255–68. https://doi.org/10.1080/00268978400101201.

[72] Pokropivny V V, Ivanovskii AL. New nanoforms of carbon and boron nitride. Russ Chem Rev 2008;77:837–73. https://doi.org/10.1070/rc2008v077n10abeh003789.

[73] Gangan A, Chakraborty B, Ramaniah LM, Banerjee S. First principles study on hydrogen storage in yttrium doped graphyne: Role of acetylene linkage in enhancing hydrogen storage. Int J Hydrogen Energy 2019;44:16735–44. https://doi.org/10.1016/j.ijhydene.2019.05.051.

[74] Xu L, Li C, Li F, Li X, Tao S. Molecular structure, electronic property and vibrational spectroscopy of C24-glycine and Gd@C24-glycine complexes. Spectrochim Acta - Part A Mol Biomol Spectrosc 2012;98:183–9. https://doi.org/10.1016/j.saa.2012.08.067.

[75] Deák P, Aradi B, Frauenheim T, Janzén E, Gali A. Accurate defect levels obtained from the HSE06 range-separated hybrid functional. Phys Rev B - Condens Matter Mater Phys 2010;81:1–4. https://doi.org/10.1103/PhysRevB.81.153203.

[76] Bajdich M, García-Mota M, Vojvodic A, Nørskov JK, Bell AT. Theoretical investigation of the activity of cobalt oxides for the electrochemical oxidation of water. J Am Chem Soc 2013;135:13521–30. https://doi.org/10.1021/ja405997s.

[77] Henkelman G, Uberuaga BP, Jónsson H. Climbing image nudged elastic band method for finding saddle points and minimum energy paths. J Chem Phys 2000;113:9901–4.

https://doi.org/10.1063/1.1329672.

[78] Mauron P, Buchter F, Friedrichs O, Remhof A, Bielmann M, Zwicky CN, et al. Stability and reversibility of LiBH4. J Phys Chem B 2008;112:906–10. https://doi.org/10.1021/jp077572r.

[79] Faye O, Szpunar JA. An Efficient Way to Suppress the Competition between Adsorption of H 2 and Desorption of n H 2 -Nb Complex from Graphene Sheet: A Promising Approach to H 2 Storage. J Phys Chem C 2018;122:28506–17. https://doi.org/10.1021/acs.jpcc.8b09498.

[80] Schur D V., Zaginaichenko SY, Savenko AF, Bogolepov VA, Anikina NS, Zolotarenko AD, et al. Experimental evaluation of total hydrogen capacity for fullerite C 60. Int J Hydrogen Energy 2011;36:1143–51. https://doi.org/10.1016/j.ijhydene.2010.06.087.

[81] Tang W, Sanville E, Henkelman G. A grid-based Bader analysis algorithm without lattice bias. J Phys Condens Matter 2009;21. https://doi.org/10.1088/0953-8984/21/8/084204.

[82] Kubas GJ. Metal-dihydrogen and s-bond coordination: the consummate extension of the Dewar-Chatt-Duncanson model for metal-olefin p bonding. vol. 635. 2001.

[83] Liu Y, Ren L, He Y, Cheng HP. Titanium-decorated graphene for high-capacity hydrogen storage studied by density functional simulations. J Phys Condens Matter 2010;22. https://doi.org/10.1088/0953-8984/22/44/445301.

[84] Sathe RY, Bae H, Lee H, Dhilip Kumar TJ. Hydrogen storage capacity of low-lying isomer of C24 functionalized with Ti. Int J Hydrogen Energy 2020;45:9936–45. https://doi.org/10.1016/j.ijhydene.2020.02.016.

# Supporting Information

# Exploring yttrium doped $C_{24}$ fullerene as a high-capacity reversible hydrogen storage material: DFT investigations

*Vikram Mahamiya[a], Alok Shukla[a*], BrahmanandaChakraborty[b,c*],*

[a]Indian Institute of Technology Bombay, Mumbai 400076, India

[b]High pressure and Synchrotron Radiation Physics Division, Bhabha Atomic Research Centre, Bombay, Mumbai, India-40085

[c]Homi Bhabha National Institute, Mumbai, India-400094

email: shukla@phy.iitb.ac.in ; brahma@barc.gov.in

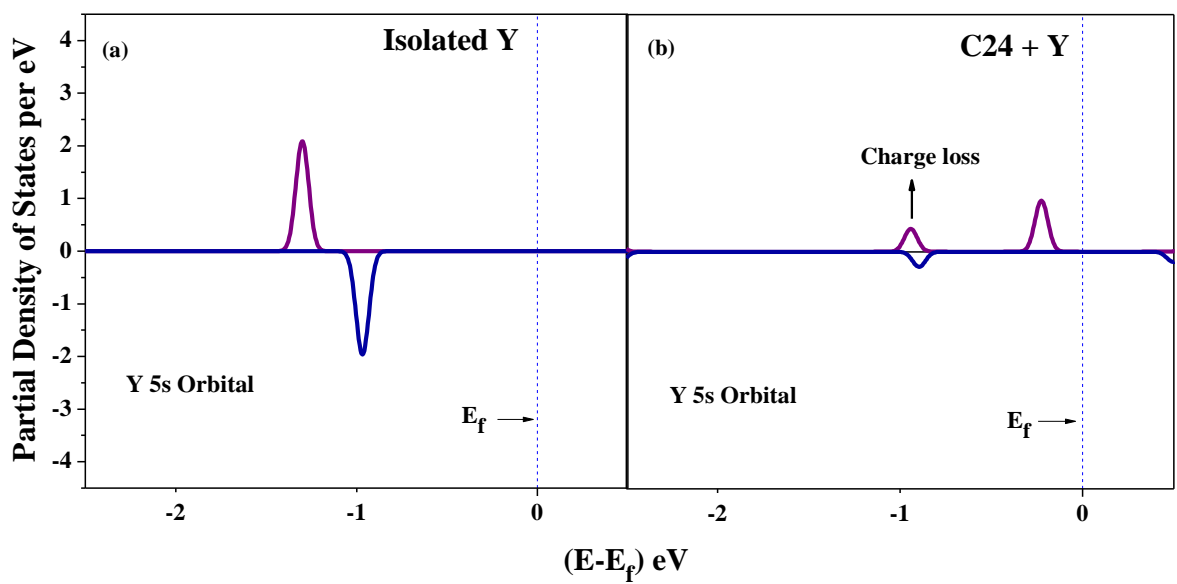

**Fig. S1** Partial density of states for (a) Y 5s orbital of isolated Y atom. (b) Y 5s orbital of $C_{24}$ + Y. Fermi energy is set at zero energy value.

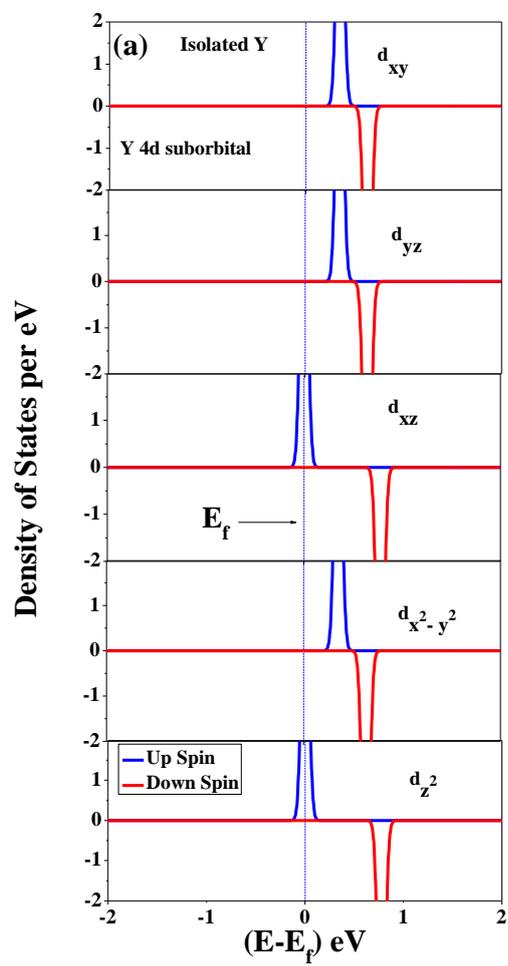

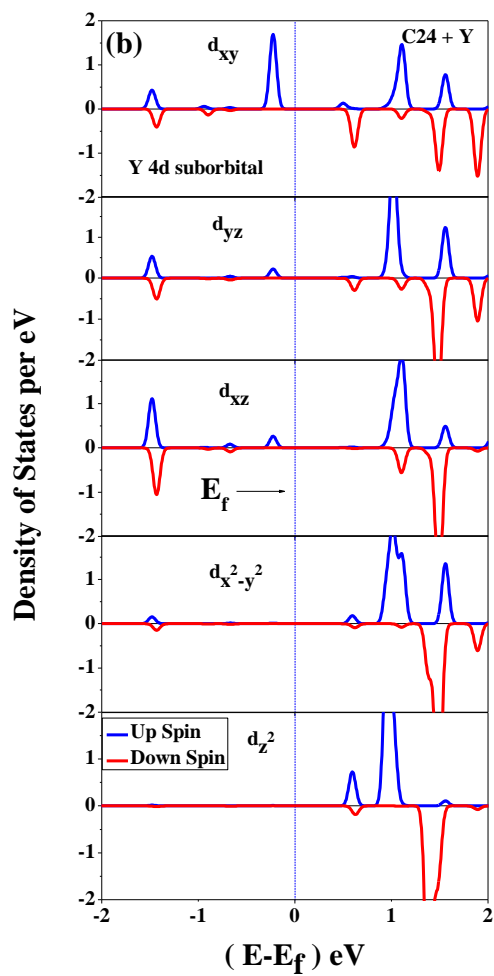

Fig. S2 Partial density of states of Y 4d sub-orbitals for (a) Isolated Y atom. (b) $C_{24}$ + Y. Fermi energy is set at zero energy value.